\documentclass[apj,twocolumn]{emulateapj}

\usepackage{float}
\usepackage{amsmath}
\usepackage{epsfig,floatflt}

\usepackage{graphicx}
\usepackage[]{subfigure}
\usepackage{color}
\definecolor{Blue}{rgb}{0.1,0.1,1.0} 
\definecolor{Magenta}{rgb}{1.0,0.1,0.5} 
\definecolor{LRed}{rgb}{0.8,0.0,0.0}

\newcommand{\nc}{\newcommand}

\nc{\be}[1]{\begin{equation}\mbox{$\label{#1}$}}
\nc{\bea}[1]{\begin{eqnarray} \mbox{$\label{#1}$}}
\nc{\Section}[2]{\section{#2}\label{#1}}
\nc{\Bibitem}[1]{\bibitem{#1}}
\nc{\Label}[1]{\label{#1}}

\nc{\eea}{\end{eqnarray}}
\nc{\ee}{\end{equation}}

\nc{\bdm}{\begin{displaymath}}
\nc{\edm}{\end{displaymath}}
\nc{\dpsty}{\displaystyle}
\nc{\bc}{\begin{center}}
\nc{\ec}{\end{center}}
\nc{\ea}{\end{array}}
\nc{\bab}{\begin{abstract}}
\nc{\eab}{\end{abstract}}
\nc{\btab}{\begin{tabular}}
\nc{\etab}{\end{tabular}}
\nc{\bit}{\begin{itemize}}
\nc{\eit}{\end{itemize}}
\nc{\ben}{\begin{enumerate}}
\nc{\een}{\end{enumerate}}
\nc{\bfig}{\begin{figure}}
\nc{\efig}{\end{figure}}

\nc{\arreq}{&\!=\!&}
\nc{\arrmi}{&\!-\!&}
\nc{\arrpl}{&\!+\!&}
\nc{\arrap}{&\!\!\!\approx\!\!\!&}
\nc{\non}{\nonumber}

\def\lsim{\; \raise0.3ex\hbox{$<$\kern-0.75em
      \raise-1.1ex\hbox{$\sim$}}\; }
\def\gsim{\; \raise0.3ex\hbox{$>$\kern-0.75em
      \raise-1.1ex\hbox{$\sim$}}\; }

\nc{\DOT}{\hspace{-0.08in}{\bf .}\hspace{0.1in}}
\nc{\Laada}{\hbox {$\sqcap$ \kern -1em $\sqcup$}}
\nc\loota{{\scriptstyle\sqcap\kern-0.55em\hbox{$\scriptstyle\sqcup$}}}
\nc\Loota{{\sqcap\kern-0.65em\hbox{$\sqcup$}}}
\nc\laada{\Loota}
\nc{\qed}{\hskip 3em \hbox{\BOX} \vskip 2ex}

\nc{\real}{{\rm I \! R}}
\nc{\Z}{{\sf Z \!\!\! Z}}
\nc{\complex}{{\rm C\!\!\! {\sf I}\,\,}}
\def\bigid{\leavevmode\hbox{\small1\kern-3.8pt\normalsize1}}
\def\id{\leavevmode\hbox{\small1\kern-3.3pt\normalsize1}}
\nc{\slask}{\!\!\!/}
\nc{\bis}{{\prime\prime}}
\nc{\pa}{\partial}
\nc{\ra}{\rangle}
\nc{\goto}{\rightarrow}
\nc{\swap}{\leftrightarrow}

\nc{\EE}[1]{ \mbox{$\cdot10^{#1}$} }
\nc{\abs}[1]{\left|#1\right|}
\nc{\at}[2]{\left.#1\right|_{#2}}
\nc{\norm}[1]{\|#1\|}
\nc{\abscut}[2]{\Abs{#1}_{\scriptscriptstyle#2}}
\nc{\vek}[1]{{\rm\bf #1}}
\nc{\integral}[2]{\int\limits_{#1}^{#2}}
\nc{\inv}[1]{\frac{1}{#1}}
\nc{\dd}[2]{{{\partial #1}\over{\partial #2}}}
\nc{\ddd}[2]{{{{\partial}^2 #1}\over{\partial {#2}^2}}}
\nc{\dddd}[3]{{{{\partial}^2 #1}\over
    {\partial #2 \partial #3}}}
\nc{\dder}[2]{{{d #1}\over{d #2}}}
\nc{\ddder}[2]{{{d^2 #1}\over{d {#2}^2}}}
\nc{\dddder}[3]{{d^2 #1}\over
    {d #2 d #3}}
\nc{\dx}[1]{d\,^{#1}x}
\nc{\dy}[1]{d\,^{#1}y}
\nc{\dz}[1]{d\,^{#1}z}
\nc{\dl}[1]{\frac{d\,^{#1}l}{(2\pi)^{#1}}}
\nc{\dk}[1]{\frac{d\,^{#1}k}{(2\pi)^{#1}}}
\nc{\dq}[1]{\frac{d\,^{#1}q}{(2\pi)^{#1}}}

\nc{\bfT}{{\bf T }}

\nc{\cA}{{\cal A}}
\nc{\cB}{{\cal B}}
\nc{\cD}{{\cal D}}
\nc{\cE}{{\cal E}}
\nc{\cG}{{\cal G}}
\nc{\cH}{{\cal H}}
\nc{\cL}{{\cal L}}
\nc{\cO}{{\cal O}}
\nc{\cT}{{\cal T}}
\nc{\cN}{{\cal N}}
\nc{\cR}{{\cal R}}

\nc{\rvac}[1]{|{\cal O}#1\rangle}
\nc{\lvac}[1]{\langle{\cal O}#1|}
\nc{\rvacb}[1]{|{\cal O}_\beta #1\rangle}
\nc{\lvacb}[1]{\langle{\cal O}_\beta #1 |}
\nc{\bb}{\bar{\beta}}
\nc{\bt}{\tilde{\beta}}
\nc{\ctH}{\tilde{\cal H}}
\nc{\chH}{\hat{\cal H}}
\nc{\al}{\alpha}
\nc{\g}{\gamma}
\nc{\Del}{\Delta}
\nc{\e}{\textrm{e}}
\nc{\eps}{\epsilon}
\nc{\lam}{\lambda}
\nc{\Om}{\Omega}
\nc{\ve}{\varepsilon}
\nc{\mn}{{\mu\nu}}
\nc{\vp}{\varphi}

\nc{\rf}[1]{(\ref{#1})}
\nc{\nn}{\nonumber \\*}
\nc{\bfB}{\bf{B}}
\nc{\bfv}{\bf{v}}
\nc{\bfx}{\bf{x}}
\nc{\bfy}{\bf{y}}
\nc{\vx}{\vec{x}}
\nc{\vy}{\vec{y}}
\nc{\oB}{\overline{B}}
\nc{\oI}{\overline{I}}
\nc{\oR}{\overline{R}}
\nc{\rar}{\rightarrow}
\nc{\ti}{\times}
\nc{\slsh}{\hskip-5pt/}
\nc{\sm}{Standard~Model~}
\nc{\MP}{M_{\rm Pl}}
\nc{\mpl}{M_{\rm Pl}}
\nc{\tp}{t_{\rm Pl}}

\nc{\pmin}{p_{\rm min}}
\nc{\pmax}{p_{\rm max}}
\nc{\fo}{f_0}
\nc{\foi}{f_{0,i}\,}
\nc{\fop}{f_0^P}
\nc{\fou}{f_0^U}

\nc{\eff}{{\rm eff}}
\nc{\MT}{M_{\rm T}}
\nc{\ML}{M_{\rm L}}
\nc{\kk}{\vek{k}}
\nc{\pp}{{\rm p}}
\nc{\pt}{\partial_t}
\nc{\half}{{1\over 2}}
\nc{\w}{\omega}
\nc{\uhat}{\hat{U}_\w}

\nc{\etal}{\mbox{\it et al.}}
\nc{\ie}{{\it i.e. }}
\nc{\eg}{{\it e.g. }}
\nc{\trh}{T_{\rm RH}}
\nc{\ad}{{a'\over a}}
\nc{\bd}{{b'\over b}}
\nc{\Rd}{{R'\over R}}
\nc{\diag}{{\textrm{diag}}}
\nc{\mato}[1]{\tilde{#1}}
\nc{\sinn}{\textrm{sinn}}
\nc{\sech}{\textrm{sech}}
\nc{\I}{\textrm{I}}
\nc{\II}{\textrm{II}}
\nc{\III}{\textrm{III}}
\nc{\vev}[1]{\langle #1 \rangle}
\nc{\hyp}{\,\; F_{1{\hskip -16pt}2}{\hskip 11pt}}
\nc{\brhom}{\overline{\rho}_M}
\nc{\brho}{\overline{\rho}}
\nc{\rhob}{\overline{\rho}}
\nc{\Pb}{\overline{P}}
\nc{\bH}{\overline{H}}
\nc{\ep}{{1+4\eps}}

\nc{\deriv}[2]{ 
\frac{\mathrm{d}#1}{\mathrm{d}#2}
}
\nc{\Mnu}{M_\nu}
\nc{\bee}{\begin{equation}}
\nc{\ene}{\end{equation}}
\nc{\hdp}{\sigma_8 (\Omega_{\rm m}/0.3)^{0.37}}
\nc{\avis}{\alpha_{vis}}
\nc{\cvis}{c^2_{vis}}
\nc{\clam}{c^2_{lam}}

\def\smiley{\hbox{\large$\bigcirc$\hspace{-.80em}%
\raise.2ex\hbox{$\cdot\cdot$}\kern-.61em    
\lower.2ex\hbox{\scriptsize$\smile$}}\ }

\def\frowney{\hbox{\large$\bigcirc$\hspace{-.80em}%
\raise.2ex\hbox{$\cdot\cdot$}\kern-.635em
\lower.2ex\hbox{\scriptsize$\frown$}}\ }

\begin{document}

\title{A self-contained guide to the CMB Gibbs sampler}
\author{Nicolaas E. Groeneboom\altaffilmark{1}}

\email{leuat@irio.co.uk}

\altaffiltext{1}{Institute of Theoretical Astrophysics, University of
  Oslo, P.O.\ Box 1029 Blindern, N-0315 Oslo, Norway}

\date{\today}

\begin{abstract} 
We present a consistent self-contained and pedagogical review of the CMB Gibbs sampler,
focusing on computational methods and code design. We provide an easy-to-use CMB Gibbs sampler
named \texttt{SLAVE} developed in C++ using object-oriented design. While
discussing why the need for a Gibbs sampler is evident and what the Gibbs
sampler can be used for in a cosmological context, we review in detail the
analytical expressions for the conditional probability densities  
and discuss the problems of
galactic foreground removal and anisotropic noise. Having
demonstrated that \texttt{SLAVE} is a working, usable CMB Gibbs
sampler, we present the algorithm for white noise level estimation. We then
give
a short guide on operating \texttt{SLAVE} before 
introducing the post-processing utilities for obtaining the best-fit
power spectrum using the 
Blackwell-Rao estimator. 
\end{abstract}

\keywords{cosmic microwave background --- cosmology: observations --- 
methods: numerical}

\maketitle

\section{Introduction}
\label{sec:introduction}
In recent years, increased resolution in the measurement of the cosmic
microwave background (CMB) have driven the need for more accurate data
analysis techniques. During the early years of CMB experiments, data
was so sparse and noise levels so high that error bars in general
overshadowed the observed signal. With the COBE experiment,
\citep{smoot:1992} posteriors were mapped out by brute force, and the
statistical methods employed were simplistic. This was
sufficient, as advanced statistical methods weren't needed for analyzing
crude data. However, all this changed with the Wilkinson Microwave
Anisotropy Probe (WMAP) experiment \citep{bennett:2003, hinshaw:2007}.
Suddenly, cosmological data became much more detailed, vastly
improving our knowledge of the universe, but also introduced new problems.
Which parts of the signal were pure CMB, and which were not?
The need for knowledge about instrumental noise, point
sources, dust emission, synchrotron radiation and other contaminations were required in order to
estimate the pure CMB signal from the data. And, how does one properly
deal with the the sky cut, the contamination from our galaxy? Even
harder, how does one maximize the probability that the resulting
signal really is the correct CMB signal? A new era of cosmological
statistics emerged.

An important event was the introduction of Bayesian statistics in
cosmological data analysis. Bayesian statistics differs from the
frequentist thought by quantizing ignorance: what one knows and
not knows are intrinsic parts of the analysis. The goal of any Bayesian
analysis is to go from the prior $P(\theta)$, or what is known about
the model, to the posterior $P(\theta|\textrm{data})$, the probability
of a model given data. This is summarized via Bayes' famous theorem:
\begin{equation}
\label{eq:bayes}
P(\theta | \textrm{data}) = \frac{P(\textrm{data} | \theta) P(\theta)}{P(\textrm{data})}.
\end{equation}
The posterior $P(\theta | \textrm{data})$ tells us something about how
well a model $\theta$ fits the data, and is obtained by multiplying
the prior $P(\theta)$, our assumption of the model, with the
likelihood $P(\textrm{data} | \theta)$, the probability that the
data fits the model.

The need for Bayesian statistics becomes evident when considering that
we only have data from one single experiment to analyze. 
Bayesian statistics merges with frequentist statistics
for large number of samples. And, in a cosmological context, we
are stuck with only one sample, a sample that we are constantly
measuring to higher accuracies. This sample is one realization of the
underlying universe model, and we are unable to obtain data from
another sample. 

In a standard Metropolis-Hastings (MH) Monte Carlo Markov chain-approach (MCMC), one
samples from the joint distribution by letting chains of ``random walkers'' transverse the
parameter space. The posterior is obtained by
calculating the normalized histogram of all the samples in the chains. The posterior
will eventually resemble the underlying joint distribution, or the
likelihood surface. This is a simple and easy-to-understand approach,
but not without drawbacks. For one, each MH step is required to test
the likelihood value of the chain at the current position in parameter
space up against a new proposed position. Many of these steps will be
rejected, 
and this is where the computational costs
usually reside. The Gibbs sampler provides something new: one
never needs to reject samples, and every move becomes accepted and
usable for building the posterior. This is done by assuming that we
have prior knowledge of the conditional distributions. These are then
sampled from, each in turn yielding accepted steps. 

However, the main motivation for introducing the CMB Gibbs sampler is
the drastically improvement in scaling. With conventional MCMC methods,
one needs to sample from the joint distribution, which results in an 
$\mathcal O(n^3)$ operation. For a white noise case, the Gibbs
sampler splits the sampling process into independently sampling from the two
conditional distributions,
which together yields a $\mathcal O(n^{1.5})$ operation. In other
words, the Gibbs sampler enables sampling the high-$\ell$ regime much
more effective than previous MCMC methods.

The problem of estimating the cosmological signal $\mathbf s$ from the
full signal by Gibbs sampling was first addressed in \citet{jewell:2004}, \citet{wandelt:2004}
and \citet{eriksen:2004b}. The ultimate goal of the Gibbs sampler is
to estimate the CMB signal $s$ from the data $d$, eliminating
noise $n$, convolution $A$, all while including the sky cut. Today, a
great number of papers have employed the Gibbs sampler since the introduction of the method
\citep{eriksen:2008a, eriksen:2008b, Dunkley:2008, cumberbatch:2009,
  groeneboom:2008b, groeneboom:2009a, eriksen:2006, rudjord:2009,
  jewell:2009, dickinson:2007, chu:2005b, dickinson:2009, larson:2007}.

\begin{figure}
\mbox{\epsfig{file=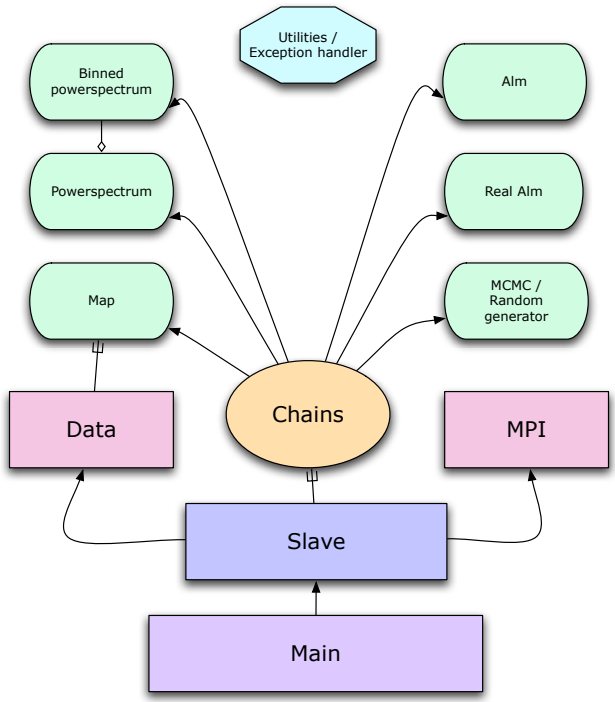, width=\linewidth,clip=}}
\caption{C++ class diagram of the \texttt{SLAVE} framework.}
\label{fig:diagram}
\end{figure}

In this paper, we review the basics of the CMB Gibbs sampler, and
provide a simple, intuitive non-parallelized CMB Gibbs software
bundle named \texttt{SLAVE}. \texttt{SLAVE} is written in C++, and employs
object-oriented design in order to simplify mathematical
implementation. The OOP design of \texttt{SLAVE} is presented in figure
\ref{fig:diagram}.
For instance, assuming $A, B$ and $C$ are
instances of the ``real alm'' class (they contain a set of real
$a_{\ell m}$s), operator overloading enables
us to directly translate the expression $A = (B+C)^{-1}$ by writing 
\begin{verbatim}
  A = (B+C).Invert();
\end{verbatim}
This yields fast code that closely resembles equations, without having
optimized too much for parallel computing, multiple data sets and
other complexities. 

\subsection{The Master algorithm}
\label{sec:master}
One method of likelihood-estimator for obtaining the best-fit
power spectrum for masked CMB data is given by the \texttt{MASTER}
algorithm \citep{hivon:2002}. While Gibbs sampling estimates the full CMB
signal $s$, the
\texttt{MASTER} method only estimates the power spectrum. This method
does not allow for variations in the estimated signal, except for the natural
variations from simulating different realizations from the same power
spectrum.  However, the master algorithm estimates the power spectrum
with cost scaling as $\mathcal O(n^3)$, which is slow for
high-$\ell$ operations.


\subsection{What do I need the CMB Gibbs sampler for?}
Often, people misunderstand the concepts behind the CMB Gibbs sampler,
and what the Gibbs sampler can be used for. In this section, we try to
explain in simple terms when you should consider employing the CMB
Gibbs sampler. 

Assume that you have a theoretical universe model $M(\theta)$, where
$\theta = \{\theta_i\}$ is a set of cosmological parameters. This
model might give rise to some additional gaussian effects in the CMB map, either as
fluctuations, altered power, anisotropic contributions, dipoles, ring structures or
whatever. You now wish to test whether existing CMB data contains
traces of your fabulous new model, and how significant those traces
are. Or maybe you are just interested in ruling out the possibility
that this model could be observed at all. 

In any case, you need to implement some sort of numerical library that
generates CMB maps based on your model. These maps will be ``pure'',
in the sense that you have complete control over its
generation process and  systematics. 
Assume that your model has 1 free parameter. You
could now loop over the 1-dimensional parameter space and calculate
the $\chi^2$ between a pure CMB signal
map and the map from your model. This would have to be done for each step
in parameter space, before obtaining the minimum. 
Even better, you could implement a Monte Carlo Markov chain framework, letting
random walkers traverse a likelihood surface, yielding posteriors. This would enable
support for a larger number of parameters, and is superior to the
slow brute force approach. 

In real-life however, things are not this simple. Data from any CMB
experiment is contaminated by noise and foregrounds, most notably our
own galaxy. This means that estimating the signal $s$ from the data 
is not trivial - one needs to ``rebuild'',
or make an assumption of what the fluctuations are 
within the sky cut and noise limits. This implies that it really
isn't possible to obtain ``the correct'' CMB map, all we can know is 
that there exist a statistical range of validity where a simulated map agrees
with the true CMB signal. Therefore, the consideration that that the estimated CMB signal $s$ is a
statistical random variable and not a fixed map should be included in
the analysis. Hence, if you have implemented the \texttt{MASTER}
method mentioned in section \ref{sec:master}, you should test your
model map against a set of realizations from the
\texttt{MASTER}-estimated signal power spectrum. 

This is where the Gibbs sampler enters the stage. As previously mentioned, the Gibbs sampler will
estimate the CMB signal given data, and not only the power spectrum. The
Gibbs sampler also ensures that every step in parameter space is always valid, so
one never needs to discard samples. And even better, each of these
independent steps provide an operation cost for
obtaining samples that are much lower than more conventional MCMC methods.
In order to test whether your model
$m$ fits the data, you therefore include the uncertainty
in data by varying the signal. For
example:
\begin{verbatim}
initialize Cl
do
  s = the CMB signal given the 
       power spectrum Cl
  m = the CMB signal of your model given 
       the estimated CMB signal s 
  Cl = the CMB power spectrum given m
  save s, m and Cl
repeat until convergence
\end{verbatim} 
In the end, you calculate the statistical properties of s, m and
Cl. Your model parameters have now been estimated, and the process included the intrinsic
uncertainties in the signal. This method is
not the most rapid - but it will always yield correct
results.

\section{The CMB Gibbs sampler}
Throughout this paper, we assume that the data can be expressed as
\begin{equation}
d = As +n
\end{equation}
where $s$ is the CMB signal, $A$ the instrument beam and $n$ uncorrelated
noise.

The \texttt{MASTER} algorithm estimates the the power spectrum $\langle \hat C_\ell \rangle$ and the
standard deviation $\Delta C_\ell$. However, this method is a approximation
to a full likelihood that can be expressed as follows:
\begin{equation}
P(C_\ell | d) = \frac{1}{\sqrt{|S+N|}} e^{-\frac{1}{2}d^T(S+N)^{-1}d}.
\label{eq:fulldist}
\end{equation}
where $S$ and $N$ are the signal and noise covariance matrices,
respectively. While it is fully possible to use MCMC-methods to sample from this
distribution, the calculation of the $(S+N)^{-1}$-matrix scales as
$n^3$, where $n$ is the size of the $n \times n$ matrix. This is therefore an extremely slow 
operation, and is not feasible for large
$\ell$s. If we demand that we
sample the sky signal $s$ as well, the joint distribution becomes
$P(C_\ell, s | d)$. This might seem unnecessary complicated, as one most of
the time doesn't need the signal $s$. But when feeding this
distribution through the Gibbs sampler - that is, calculating the
conditional distributions $P(C_\ell | s,d) $ and $P(s | C_\ell, d)$,
we find that sampling from both are computationally faster 
than sampling from the full distribution in equation \ref{eq:fulldist}. The
derivations of the conditional distributions are presented in section \ref{sec:methods}.

\subsection{Review of the Metropolis-Hastings algorithm}
The Gibbs sampler is a special case of the Metropolis-Hastings
algorithm. We therefore review the basics of Monte Carlo Markov (MCMC)
chain methods.
The Metropolis-Hastings algorithm is a MCMC
method for sampling directly from a probability distribution. This is
done by letting ``random walkers'' transverse a parameter space,
guided by the likelihood function, the probability that the data fits
the model for the given parameter configuration. If a proposal step yields a
likelihood greater than the current likelihood, then random walker
accepts the step immediately. If the likelihood is less, then the
walker will with a certain probability step ``down'' the likelihood
surface. Eventually, the histogram of all the random walkers will
converge to the posterior, the full underlying distribution.

Assume you have a model with $n$ parameters, $\theta = \{\theta_k\}$ and you wish to map out a joint distribution from
$P(\theta)$. Usually, one calculates the ratio $R$ between the posteriors at the
two steps $P(\theta^{i+1})$ and $P(\theta^{i})$, such that
\begin{equation}
  \label{eq:asymmprop}
  R = \frac{P(\theta^{i+1})}{P(\theta^{i})} \cdot 
   \frac{T(\theta^{i} | \theta^{i+1})}{T(\theta^{i+1} | \theta^{i})}
\end{equation}
where $T(\theta^{i} | \theta^{i+1})$ is the proposal distribution for
going left or right. If the proposal
distribution is symmetric (i.e. the probability of going left-right is
equal for all $\theta_k$), then 
$T(\theta^{i} | \theta^{i+1}) = T(\theta^{i+1} | \theta^{i})$ such that:
\begin{equation}
  R = \frac{P(\theta^{i+1})}{P(\theta^{i})}
\end{equation}
The MH acceptance rule now states: if $R$ is larger than 1, accepted the step unconditionally.
If $R>1$, then accept the step if a random uniform variable $x =
U(0,1) < R$.

\subsection{Review of the Gibbs algorithm}
\label{sec:method_gibbs}
Assume you have a model with two parameters, $\theta_1$ and
$\theta_2$, and you wish to map out a joint distribution from
$P(\theta_1, \theta_2)$. Now, also presume that you have prior
knowledge of the conditional distributions, $P(\theta_1 | \theta_2)$
and $P(\theta_2 | \theta_1)$. A general proposal density is not
necessary symmetric, and one must therefore consider the asymmetric
proposal term as described in equation \ref{eq:asymmprop}.  However,
we now define the proposal density $T$ for $\theta_2$ to be the
conditional distributions:
\begin{equation}
  T(\theta^{i+1}_1, \theta_2^{i+1} | \theta^i_1, \theta^i_2) = 
\delta(\theta^{i+1}_1 - \theta^{i}_1) P(\theta^{i+1}_2 | \theta^i_1).
\label{eq:prop}
\end{equation}
In words, the proposal is only considered when $\theta^{i+1}_1 =
\theta^{i}_1$, which means that $\theta_1$ is fixed while
$\theta_2$ can vary.  If so, the acceptance is then given as the
conditional distribution $P(\theta^{i+1}_2 | \theta^i_1)$, which we
must have prior knowledge of. The reason for choosing such a proposal
density becomes clear when investigating the Metropolis Hastings
acceptance rate:
\begin{equation}
   R = \frac{P(\theta^{i+1}_2 , \theta^{i+1}_1 )}{P(\theta^{i}_2 , \theta^i_1)} \cdot 
   \frac{T(\theta^{i}_1,\theta^{i}_2 | \theta^{i+1}_1,\theta^{i+1}_2 )}
{T(\theta^{i+1}_1,\theta^{i+1}_2 | \theta^{i}_1,\theta^{i}_2)}
\end{equation}
Using the conditional sampling proposal (\ref{eq:prop}) one obtains
\begin{equation}
  R = \frac{P(\theta^{i+1}_2 | \theta^{i+1}_1 ) P(\theta^{i+1}_1)}{P(\theta^{i}_2 | \theta^i_1)P(\theta^i_1)} \cdot 
  \frac{P(\theta^i_2 | \theta^{i+1}_1)}{P(\theta_2^{i+1} | \theta_1^i)}
  \frac{\delta}{\delta}
\end{equation}
We now enforce the delta-function such that $\theta_1^{i+1} =
\theta_1^i$. This sampling from the conditional distributions is the
crucial step in the Gibbs sampler, 
such that all terms cancel out: 
\begin{equation}
  R = 1.
\end{equation}
This implies that all steps are valid, and none are ever rejected. Hence one alternates between
sampling from the known conditional distributions, where each step is
independently accepted and can be performed as many times as needed.

\begin{figure}[h]
\label{sampling}
\begin{center}
 \input{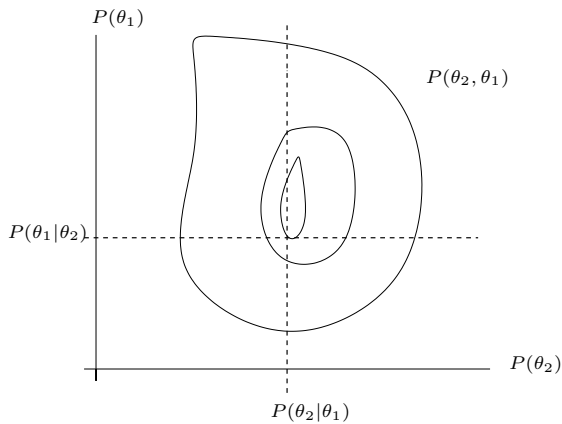}
\end{center}
\caption{Conditional sampling implies alternating between sampling
  from $P(\theta_1 | \theta_2)$ and $P(\theta_2 | \theta_1)$, fixing the
  other parameter.}
\end{figure}

\section{The conditional distributions}
\label{sec:methods}
In section \ref{sec:method_gibbs}, it was explained how the Gibbs
sampler requires previous knowledge about the underlying conditional
distributions. The CMB Gibbs sampler will alternate between sampling
power spectra $C_\ell$ and CMB signal $s$, where each proposed step will always
be valid. In order to enable sampling from the joint
distribution, we therefore need to derive the analytical properties of
the conditional distributions:
\begin{equation}
  P(C_\ell |s, d)  \hspace{7mm} \textrm{and}  \hspace{7mm}  P(s | C_\ell, d).
\end{equation}
The derivations described here were first presented in \citet{jewell:2004}, \citet{wandelt:2004}
and \citet{eriksen:2004b}.The full, joint distribution is expressed as
\begin{eqnarray}
  P(C_\ell, s | d) & \propto & P(d|C_\ell, s)P(C_\ell, s) \\
                  & =        & P(d|C_\ell,s )P(s | C_\ell)P(C_\ell)
\end{eqnarray}
where $P(C_\ell)$ is a prior on $C_\ell$, typically chosen to be flat. 
The first term, $-2\ln P(d|C_\ell,s )$, is nothing but the
$\chi^2$. The $\chi^2$ measures the goodness-of-fit between model and
data, leaving only fluctuations in noise. As $n = d-s$ is distributed
accordingly to a Gaussian with mean 0 and variance $N$, we find that
\begin{equation}
  P(d | C_\ell, s) \propto 
e^{-\frac{1}{2}(d-s)^tN^{-1}(d-s)}.
\label{eq:chisq}
\end{equation}

As we now assume that the signal $s$ is known and fixed, the data
$d$ becomes redundant and 
$P(C_\ell | s,d ) = P(C_\ell| s) \propto P(s | C_\ell)$. We therefore first need
to obtain an expression for $P(C_\ell | s, d)$.

\subsection{Deriving $P(C_\ell | s, d)$}

Assuming that the CMB map consists of Gaussian fluctuations, we can
express the conditional probability density for a power spectrum $C_\ell$  given a
sky signal $s$ as follows:
\begin{equation}
  P(C_\ell | s, d) = \frac{e^{-\frac{1}{2}s^TC^{-1}s}}{\sqrt{|C|}}
\label{eq:pscl}
\end{equation}
where $C=C(C_\ell)$ is the covariance matrix. We now perform a transformation to
spherical harmonics space, where $s = \sum_{\ell m}a_{\ell m}Y_{lm}$
and $C_{ij} = \sum_i \sum_j Y_{\ell' m'}^iC_{\ell 'm',\ell m}Y_{\ell m}^j$. Then
equation (\ref{eq:pscl}) transforms to
\begin{equation}
  s^TC^{-1}s = \sum_{\ell m}\sum_{\ell' m'}  a_{\ell m}^*Y_{\ell m}^* Y_{\ell' m'}C^{-1} Y_{\ell m}^* Y_{\ell' m'} a_{\ell' m'}.
\end{equation}
As the spherical harmonics are orthogonal, they all cancel out and leave delta functions for $\delta_{\ell \ell'}
\delta_{m m'}$ such that
\begin{equation}
  s^TC^{-1}s = \sum_{\ell m}a_{\ell m}^*C_\ell^{-1} a_{\ell m} = 
\sum_{\ell m}a_{\ell m}^* \frac{1}{C_\ell} a_{\ell m}.
\end{equation}
We now define a power spectrum $\sigma_\ell = \frac{1}{2\ell +1}\sum_m |a_{\ell m}|^2$ such that
\begin{equation}
  s^TC^{-1}s = \sum_{\ell} (2\ell+1) \frac{\sigma_l}{C_\ell}. 
\end{equation}

Similarly, the determinant is given as the product of the diagonal matrix $C$, which for each
$\l$ has $2\ell +1$ values of $C_\ell$. The determinant is thus $|C| = \prod_\ell C_\ell^{2\ell +1}$.
Expression (\ref{eq:pscl}) can now be written as
\begin{equation}
  P(C_\ell |s) 
= \prod_\ell \frac{e^{-\frac{(2\ell +1)}{2} \frac{\sigma_\ell}{C_\ell}}}{\sqrt{C_\ell^{2\ell +1}}}
\label{eq:invgamma}
\end{equation}
which by definition means that the $C\ell$'s are distributed as an inverse
Gamma function. In the computational section, we will discuss how to
draw random variables from this distribution.

\subsection{Deriving $P(s | C_\ell, d)$}
Again, we begin with the full, joint distribution:
\begin{equation}
  P(C_\ell, s | d) \propto P(d|C_\ell, s)P(C_\ell| s).
\label{eq:fulls}
\end{equation}
We now know from equation \ref{eq:invgamma} and \ref{eq:chisq}  that the joint distribution can be expressed as
\begin{equation}
  P(C_\ell, s | d) \propto 
e^{-\frac{1}{2}(d-s)^tN^{-1}(d-s)}
\prod_\ell \frac{e^{-\frac{2\ell+1}{2}\frac{\sigma_\ell}{C_\ell}
  }}{C_\ell^{\frac{2\ell+1}{2}}}
\label{eq:fullpost}
\end{equation}
omitting the prior $P(C_\ell)$. Again, note that it would be
nearly impossible to sample directly from the full distribution.
We now investigate what happens with equation \ref{eq:fullpost} when
$C_\ell$ becomes a fixed quantity. As the $C_\ell$s in the denominator
vanishes, we use equation \ref{eq:pscl} to obtain
\begin{equation}
  P(s | C_\ell, d) \propto e^{-\frac{1}{2}(d-s)^TN^{-1}(d-s)} e^{-\frac{1}{2}s^TC^{-1}s}.
\label{eq:scldbefore}
\end{equation}

We now introduce a residual variable $r = d - s$, 
such that $r$ roughly consist of noise. As noise was uncorrelated,
we can expect that $r$ follows a Gaussian distribution with zero mean and $N$ variance. 
Also, if $s$ is known, then $C_\ell$ is redundant. 
We complete the square, and introduce $\hat s = (S^{-1} + N^{-1})^{-1}N^{-1}d$. Equation (\ref{eq:scldbefore}) can now
be rewritten as 
\begin{equation}
  P(s | C_\ell, d) \propto e^{-\frac{1}{2}(s-\hat s)^T(C^{-1} + N^{-1})(s-\hat s)}. 
\label{eq:scomplete}
\end{equation}
Hence $P(s | C_\ell, d)$ is a Gaussian distribution with mean $\hat s$
and covariance $(C^{-1} + N^{-1})^{-1}$.  In the computational
section, we will discuss how to draw random variables from this
distribution.

\section{Numerical implementation}
In its utter simplicity, the mechanics of the Gibbs sampler can be
summarized as follows:
\begin{verbatim}
 load data
 initialize s and cl
 loop number of chains
   s = generate from p(s | cl, d)
   cl = generate from p(cl | s, d)
   save s and cl
 end loop
\end{verbatim}
We now present the computational methods for drawing from $P(s |
C_\ell, d)$ and $P(C_\ell| s, d)$.

\subsection{$P(C_\ell| s, d)$}
We show that equation \ref{eq:invgamma} is an inverse
Gamma distribution. A general gamma-distribution is proportional to
\begin{equation}
  P_\Gamma(x; k,\theta) \propto x^{k-1}e^{-\frac{x}{\theta}}.
\end{equation}
Equation \ref{eq:invgamma} can be expressed as 
\begin{equation}
  P(C_\ell |s)  = C_\ell^{-\frac{2l +1}{2}}e^{-\beta/C\ell}
\end{equation}
where $\beta = \frac{2l +1}{2}\sigma_i$. If we now perform
a substitution $y = 1/C_\ell$, we see that 
\begin{equation}
  P(y|s)  = y^{\frac{2l +1}{2}}e^{-\beta y} \cdot  y^{-2}
\end{equation}
where the last term is the Jacobian. Hence
\begin{equation}
  P(y|s)  = y^{\frac{2l -1}{2} -1}e^{-\beta y} 
\end{equation}
which is a gamma-distribution for $k = \frac{2l -1}{2}$. We now show
that this particular distribution also happens to be a special case of 
the $\chi^2$ distribution:
\begin{equation}
  \chi(x; k) = x^{k'/2-1}e^{-\frac{x}{2}}.
\end{equation}
Letting $z = 2\beta y$ and ignoring the constants, we find that
\begin{equation}
  P(z|s)  = z^{k -1}e^{-z/2} 
\end{equation}
such that if $k' = 2k = 2l-1$, $z$ is distributed according to a
$\chi^2$ distribution with $2l-1$ degrees of freedom. A random
variable following such a distribution can be drawn as follows:
\begin{equation}
  z_\chi = \sum_{i=0}^{2l-1}|N_i(0,1)|^2
\end{equation}
where $N_i(0,1)$ are random Gaussian variables with mean $0$ and variance
$1$. Since $z = 2\beta y = 2\beta / C_\ell$, we find that
\begin{equation}
  C_\ell = (2l+1) \sigma_i /z_\chi.
\end{equation}
Numerically, one can implement this as 
\begin{verbatim}
 for each l
  z = 0
  for i=0 to 2l-1
    z = z+ rand_gauss()^2
  end
  C(l) = (2l+1)*sigma(l)/z
end
\end{verbatim}
An example of this method can be found in the \texttt{SLAVE} libraries, within
class ``powerspectrum'' method ``draw\_gamma''.

\subsection{$P(s | C_\ell, d)$}
From equation \ref{eq:scomplete}, it is easy to see that $P(s |
C_\ell, d)$ is a Gaussian distribution with mean $\hat s$ and variance
$(C^{-1} + N^{-1})^{-1}$. Instead of deriving a method for drawing a random
variable from this distribution, we present the solution and show that
this solution indeed has the necessary properties \citep{jewell:2004}. Let
\begin{equation}
  s = (C^{-1} + N^{-1})^{-1}(N^{-1}d + N^{-\frac{1}{2}}\omega_1 + C^{-\frac{1}{2}} \omega_2)
\label{eq:draws}
\end{equation}
where $\omega_1$ and $\omega_2$ are independent, random $N(0,1)$
variables. We now show that the random variable $s$ indeed has
mean $\hat s$ and variance $(C^{-1} + N^{-1})$. First, 
\begin{equation}
  \langle s \rangle = (C^{-1} + N^{-1})^{-1}(N^{-1}\langle d\rangle +
  N^{-\frac{1}{2}}\langle \omega_1 \rangle + C^{-\frac{1}{2}} \langle
  \omega_2\rangle ).
\end{equation}
As $\langle \omega_1\rangle = \langle \omega_2\rangle = 0$, 
\begin{equation}
  \langle s \rangle = (C^{-1} + N^{-1})^{-1}N^{-1}\langle d\rangle  =
  \hat s
\end{equation}
by definition. 

The covariance is then
\begin{equation}
  \langle (s-\hat s)(s - \hat s)^T \rangle.
\end{equation}
Note that in the term $s-\hat s$, we have $(C^{-1} + N^{-1})^{-1}(N^{-1}d - N^{-1}d)=0$, so we are only left
with the terms with the random variables $\omega$:
\begin{eqnarray*}
  \langle (s-\hat s)(s - \hat s)^T \rangle &= & 
(C^{-1} + N^{-1})^{-2} \cdot \\
 & & 
  \langle (N^{-\frac{1}{2}} \omega_1 + C^{-\frac{1}{2}}\omega_2  )(
  \omega_1^TN^{-\frac{T}{2}}  +  \omega_2^T C^{-\frac{T}{2}}) \rangle
\end{eqnarray*}
But, as $\omega_1$ and $\omega_2$ are independently drawn from a
$N(0,1)$ distribution, then
$\langle \omega_i \omega_j \rangle = \delta_{ij}I$, and we end up with
\begin{equation}
  \langle (s-\hat s)(s - \hat s)^T \rangle = (C^{-1} + N^{-1})^{-1}
\end{equation}
which shows that a random variable drawn using equation \ref{eq:draws}
has the desired properties of being drawn from $P(s|C_\ell, d)$.

Having implemented a ``real alm'' class in \texttt{SLAVE} with operator
overloading, it is possible to directly translate equation \ref{eq:draws} into
code:
\begin{verbatim}
  omega1.gaussian_draw(0, 1, rng);
  omega2.gaussian_draw(0, 1, rng);
  calculate_CNI();
  S = CNI* (NI*D + NI.square_root()*omega1 
     + CI.square_root()*omega2);
\end{verbatim}
where the code has been slightly optimized: both $C^{-1}$, $N^{-1}$ and
$(C^{-1} + N^{-1})^{-1}$ has been pre-calculated for
efficiency. Note that this is only possible to do when
assuming full-sky coverage with constant RMS noise. If the noise
isn't constant on the sky, then $N$ is a dense off-diagonal
matrix, nearly impossible to calculate directly for large $\ell$. 
However, it is still possible to perform the calculation in pixel
space, but this requires that we assume $N$ to be an
operator instead of a matrix. We will address this issue in section \ref{sec:skycut}.

We have now presented the main simplified Gibbs-steps for calculating
$P(s|C_\ell, d)$ and $P(C_\ell | s, d)$, without convolution, uniform
noise and no sky cut. Sampling from these two
distributions is then done alternating between the two Gibbs steps,
and the chain output - $s$ and $C_\ell$ - are saved to disk during
each step. 

We now investigate the behavior of these fields, as each
have special properties. 
\subsection{Field properties}
\begin{figure}
\mbox{\epsfig{file=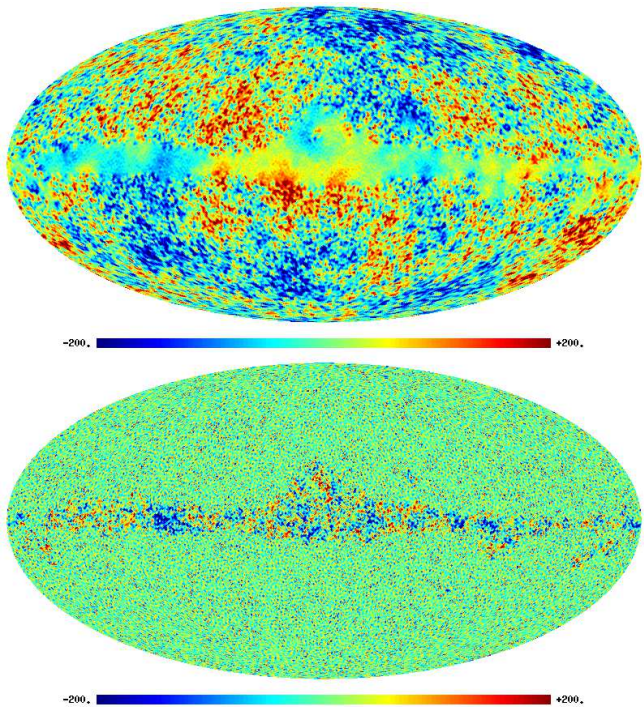, width=\linewidth,clip=}}
\caption{The two maps that together compose the full signal: the
  fluctuation map (bottom) and the Wiener filter (top). Note that
  within the sky cut, the
  Wiener filter successfully estimates the large-scale structures
  while the fluctuation map produces random
  small-scale fluctuations. }
\label{fig:wiener}
\end{figure}

Equation \ref{eq:draws} can be broken into two separate parts:
the Wiener filter $(C^{-1} + N^{-1})^{-1}(N^{-1}d)$ and the
fluctuation map $(C^{-1} + N^{-1})^{-1} (N^{-\frac{1}{2}}\omega_1 +
C^{-\frac{1}{2}} \omega_2)$. In figure \ref{fig:wiener}, each of these
maps are depicted. The Wiener filter map determines the fluctuations
outside the sky cut - where they are heavily constrained by the known
data, given cosmic variance and noise. However, within the sky cut,
large-scale fluctuations are possible to pin down statistically while
small-scales are repressed. The fluctuation map determines the
small-scale fluctuations within the unknown sky cut, and are constrained by cosmic
variance and noise effects. Outside the sky cut, the fluctuation map
is constrained by the data, yielding very low small-scale
fluctuations. The sum of these two parts make up the full
CMB signal sample.

\subsection{Verifying the sample signal: the $\chi^2$ test}
\label{sec:chisq}
When the signal is being sampled, it is vital to check that the input
parameters/data maps are correctly set up. For instance, if you use \texttt{SLAVE}
to start a large job, say, estimating the CMB signal $s$ for a
$n_{pix}=512$ map, it can be very frustrating when realizing 
that one of the input parameters were incorrect, for instance beam
convolution or noise RMS. 
The software will continue to run without errors, but the
resulting output files will be incorrect. We therefore adopt a
simple and useful method for verifying that the estimated CMB signal
$s$ for each Gibbs step really is close to what one would expect.

The trick lies with the noise. As $d = As + n$, then $n =
d-As$. Uniform white noise is assumed to be $N(0,\sigma^2_{\textrm{RMS}})$-distributed,
so 
\begin{equation}
N(0,1) \sim \frac{d-As}{\sigma_{\textrm{RMS}}}.
\end{equation}
A $\chi^2$ distribution is nothing but a sum of squared Gaussian
distributions. Hence
\begin{equation}
\chi_{n_{\textrm{pix}}}^2 \sim \sum_{n_{\textrm{pix}}}\large(\frac{d-As}{\sigma_{\textrm{RMS}}}\large)^2
\end{equation}
and the $\chi^2$ should be close to the number of pixels in
the map plus minus $\sqrt{2n}$. Usually, when an incorrect parameter
is used, the $\chi^2$ comes out far away from the expected value. 

Calculating the $\chi^2$ is not particularly time-consuming, but it
has other uses as well: the $\chi^2$ is used in the estimation of
noise, as presented in section \ref{sec:noiseestimation}.

\subsection{Convolution}
A thing we did not address in the previous section was the inclusion
of the instrumental beam convolution $A$. Including this in equation
\ref{eq:draws}, we obtain
\begin{equation}
  (C^{-1} + A^TN^{-1}A)s = AN^{-1}d + AN^{-\frac{1}{2}}\omega_1 + C^{-\frac{1}{2}} \omega_2.
\end{equation}
In \texttt{SLAVE}, the beam is loaded directly from a fits file, or generated as a
Gaussian beam given a full width half-maximum (FWHM) range. The beam
is then multiplied with the corresponding pixel window, and stored in
the $a_{\ell m}$-object $A$ throughout the code.

\subsection{The sky cut}
\label{sec:skycut}
Until now, we have only assumed full-sky data
sets contaminated by constant noise. However, in order to be able to investigate
real data, we need to take into account both the foreground galaxy and
anisotropic noise. The galaxy contributes to almost 20\% of the WMAP data, and needs to be
removed with a mask. This means that the usable pars of the maps becomes
anisotropic, giving rise to correlations in the spherical harmonics
$a_{\ell m}$s. In other words, all the previously diagonal and well-behaved matrices
now have off-diagonal elements, which for large $\ell_{\textrm{max}}$ is an
impossible feat to perform for dense matrices. 

One way to get around these problems is to perform the calculations
containing the sky cut mask in pixel space. This means that every time
one needs to take into account the sky cut, one transforms from
harmonic to pixel space, performs the operation including the sky cut
before transforming back to harmonic space. While this operation in
itself is trivial, equation \ref{eq:draws} provides a few other problems:
\begin{equation}
  (C^{-1} + A^TN^{-1}A)s = AN^{-1}d + AN^{-\frac{1}{2}}\omega_1 +
  C^{-\frac{1}{2}} \omega_2.
\label{eq:draws2}
\end{equation}
The right-hand side can easily be calculated, letting $N^{-1}$
be an operator acting on $d$ and $\omega_1$, switching from
spherical harmonics to pixel space and back. However, the left-hand
side is troublesome - one cannot solve this equation
explicitly. First, we need to rewrite \ref{eq:draws2} a bit:
\begin{eqnarray}
  (1 +
  C^{\frac{1}{2}}A^TN^{-1}AC^{\frac{1}{2}})(C^{-\frac{1}{2}}s) = \\
  C^{\frac{1}{2}}AN^{-1}d + C^{\frac{1}{2}}AN^{-\frac{1}{2}}\omega_1 +
  \omega_2 = b
\label{eq:draws3}
\end{eqnarray} 
The first thing one should note about equation \ref{eq:draws3} is that
the left-hand term is proportional to $(1 + S/N)$, where the diagonal
parts are just the signal-to-noise ratios of the corresponding
mode. Another nice feature about this form is that the variance of the
signal is kept constant, that is, $\textrm{Var}(s)
\sim \ell^{-2}$, but $\textrm{Var}(C^{-1/2}s) \sim I$. Hence we obtain better
numerical stability. In order to solve the equation $(1+S/N)x = b$, we implement a
direct-from-textbook Conjugate Gradient (CG) algorithm presented on
page 40 in \cite{Shewchuk}. The code looks like this: 
\begin{verbatim}
  b = L*( A*NI(D) + A*NI(map_work2,true)) + omega2;
  MI = setup_preconditioner();
  x = mult_by_A(x);
  r =  b - x;
  d = MI*r;
  r0 = r.norm_L1(r);
  do {
    Ad = mult_by_A(d);
    alpha = r.dot(MI*r) / (d.dot(Ad));
    x = x + d*alpha;
    rn = r - Ad*alpha;
    beta = rn.dot(MI*rn) / (r.dot(MI*r));
    d = MI * rn + d*beta;
    r = rn;
    norm = r.norm_L1(r);
  } 
  while (norm>r0*epsilon);
  S = L*x;
\end{verbatim}
C++ enables the CG algorithm to be translated almost directly from
mathematical syntax to code. Here, the sky cut mask is taken into account
in the $NI$-method - one only needs the mask when multiplying with the inverse
noise matrix. The only other ``initial condition'' is the
preconditioner. The 
preconditioner cannot affect the result, that is, it has nothing to do
with the estimated signal $s$. The preconditioner only affects the
number of iterations needed for the equation $Ax = b$ to be
solved, and corresponds to a ``best guess'' of $A$.
 Without going into details, the standard preconditioner in
\texttt{SLAVE} is proportional to $(1+S/n)$, but there exists many
other suggestions for better pre-conditioners, yielding quicker
convergence. See \cite{eriksen:2004} or \cite{smith:2007} for more examples.

When the CG search has completed, the
signal $S$ has been obtained, including the sky cut and anisotropic
noise. 

\subsection{Low signal-to-noise regime}
A final thing we need to take into account is the low signal-to-noise
regime. When the noise starts dominating the signal, the estimated
$s$ will fluctuate wildly on small scales. In addition, the
deconvolution will add to this effect, blowing up noise to extreme
values. In itself, this isn't a bad thing as we really cannot say
exactly what is going in this regime, but it will affect the overall
correlations between chains. In order to reduce this effect, we
present a simple way to bin multipoles together on large l, reducing
noise variance.

Let $N_\ell = \sigma^2_{\textrm{RMS}}4\pi/n_{pix}$ be the noise RMS in harmonic
space. The variance is then given as 
\begin{equation}
  Var(N_\ell) = \frac{2}{2l+1}N_l^2.
\end{equation}
For a single binned set with $n$ multipoles ranging from $\ell_{\textrm{low}}$ to
$\ell_{\textrm{high}}$, the average value of the power spectrum is given as 
\begin{equation}
 D_\ell = \frac{1}{n}\sum_{\ell_{\textrm{low}}}^{\ell_{\textrm{high}}}C_\ell.
\end{equation}
Similarly for the noise power spectrum, 
\begin{equation}
 N_b = \frac{1}{n}\sum_{\ell_{\textrm{low}}}^{\ell_{\textrm{high}}}N_\ell.
\end{equation}
Thus, the variance of the noise is given as 
\begin{equation}
  \sigma_N^2 = \textrm{Var}(N_b) =
  \frac{1}{n^2}\sum_{\ell_{\textrm{low}}}^{\ell_{\textrm{high}}} \textrm{Var}(N_\ell).
\end{equation}
Obviously, $\sigma_N$ is reduced as the number of multipoles in the
bin $n$ is increased. We now select bins such that the noise
variance in a single bin is always less than three times the value of
the angular power spectrum, or $\sigma_n<3 D_\ell$.

\begin{figure}[htb]
\mbox{\epsfig{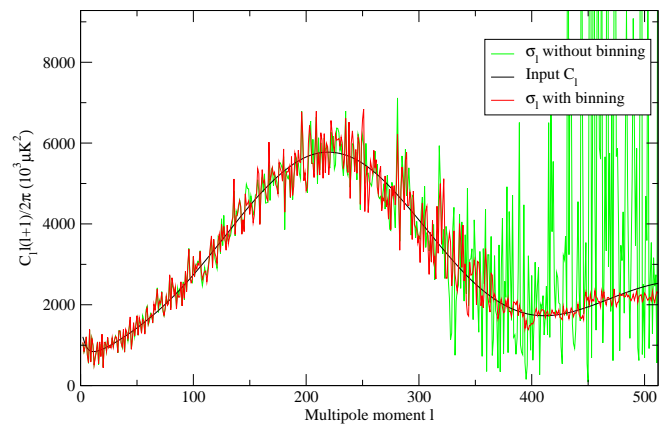}}
\caption{Examples of two estimated $\sigma_\ell$ without binning
  (green) and with binning (red). If the $C_\ell$s are produced from
  the binned $\sigma_\ell$s, the fluctuations in the low S/N-regime become
  less volatile. The input power spectrum is depicted in black.}
\label{fig:plotbins}
\end{figure}
The only affected part of the code is where one determines $P(C_\ell |
s,d)$. Instead of generating a power spectrum $C_\ell$ given a set of
$\sigma_l$, the calculation is now performed via a binning class that
calculates the binned power spectrum $C_b$. That is,
\begin{equation}
  P(C_b |\sigma) = \prod_{\ell_{\textrm{low}}}^{\ell_{\textrm{high}}} (
  \frac{e^{-{\frac{2l+1}{2}{\frac{\sigma_\ell}{C_b}}}}}{C_b^{{\frac{2\ell+1}{2}}}}
  ).
\end{equation}
Absorbing the product into the exponential, this becomes
\begin{equation}
  P(C_b |\sigma) = \frac{e^{-\frac{1}{2C_b}\sum_\ell(2l+1)\sigma_\ell}}{C_b^{\frac{1}{2}\sum_\ell(2\ell+1)}}.
\end{equation}
We now sample the signal with flat bins in $\ell(\ell+1)/(2\pi)$, not
in $\ell$. 

\section{Generalizing the model: Noise estimation}

\label{sec:noiseestimation}
In this section, we give a direct example of how one could extend the data
model to the \texttt{SLAVE} Gibbs sampler. We derive the
necessary conditional distribution, explain how this was integrated, 
and present some results from \cite{groeneboom:2009a}, where a full
analysis of the noise levels in the WMAP data was performed using the
\texttt{SLAVE} framework. 

Traditionally, the noise properties used in the Gibbs sampler
\citep[e.g.,][]{eriksen:2004} have been assumed known to infinite
precision. In this section, however, we relax this assumption, and
introduce a new free parameter, $\alpha$, that scales the fiducial
noise covariance matrix, $N^{\textrm{fid}}$, such that $N = \alpha
N^{\textrm{fid}}$. Thus, if there is no deviation between the assumed and
real noise levels, then $\alpha$ should equal 1. The full analysis of
the 5-yr WMAP data was presented in \cite{groeneboom:2009a}, with interesting results. 
For the foreground-reduced 5-year WMAP sky
maps, we find that the posterior means typically range between
$\alpha=1.005\pm0.001$ and $\alpha=1.010\pm0.001$ depending on
differencing assembly, indicating that the noise level of these
maps are underestimated by 0.5-1.0\%. The same problem is not
observed for the uncorrected WMAP sky maps.

The full joint posterior, $P(s,C_\ell, \alpha \,|\, d)$, now includes
the amplitude $\alpha$. We can rewrite this as follows:
\begin{equation}
\label{eq:postnoise}
  P(s,C_\ell, \alpha \,|\,d) = P(d \,|\,s, \alpha) \cdot P(s, C_\ell) \cdot P(\alpha)
\end{equation}
where the first term is the likelihood,
\begin{equation}
  P(d \,| \,s, \alpha) = \frac{e^{-\frac{1}{2}(d-s)(\alpha N)^{-1}(d-s)
    }}{\sqrt{|\alpha N|}},
\end{equation}
the second term is a CMB prior, and the third term is a prior on
$\alpha$. Note that the latter two are independent, given that
these describe two a-priori independent objects.  In this paper, we
adopt a Gaussian prior centered on unity on $\alpha$, $P(\alpha) \sim
N(1,\sigma_\alpha^2)$. Typically, we choose a very loose prior,
such that the posterior is completely data-driven.

The conditional distribution for $\alpha$ can now be expressed as
\begin{equation}
  P(\alpha \,|\, s,C_\ell, d) \propto
  \frac{e^{-\frac{\beta}{2\alpha}}}{\alpha^{n/2}}  
 \cdot P(\alpha)
\end{equation}
where $n=N_{\textrm{pix}}$ and $\beta = (d-s)N^{-1}(d-s)$ is the
$\chi^2$. (Note that the $\chi^2$ is already calculated within
  the Gibbs sampler, as it is used to validate that the input noise
  maps and beams are within a correct range for each Gibbs
  iteration. Sampling from this distribution within the Gibbs sampler
  represent therefore a completely negligible extra computational
  cost.) For the Gaussian prior with unity mean and standard
deviation $\sigma_\alpha$, we find that
\begin{equation}
  P( \alpha \,|\, s, C_\ell, d ) \propto
  \frac{e^{-\frac{1}{2}(\frac{\beta}{\alpha} + \frac{ (\alpha-1)^2 }{
        \sigma_\alpha^2 })}}{\alpha^{n/2}}  
\label{eq:finally}
\end{equation}
\begin{figure}
\mbox{\epsfig{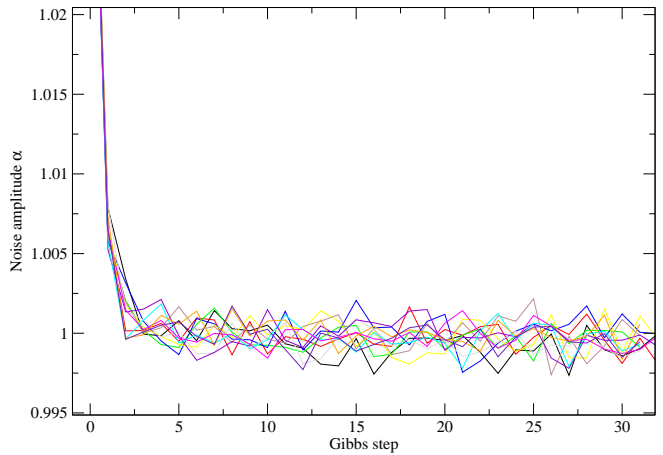}}
\caption{Even when assuming a large initial value, the noise amplitude $\alpha$
  will quickly converge to the correct value. }
\label{fig:alpha_simulated}
\end{figure}

For large degrees of freedom, $n$, the inverse gamma function
converges to a Gaussian distribution with mean $\mu = b/(k+1)$, where
we have defined $k = n_{\textrm{pix}}/2 -1$, and variance $\sigma^2 =
b^2/((k-1)(k-1)(k-2))$. A good approximation is therefore letting
$\alpha_{i+1}$ be drawn from a product of two Gaussian distributions,
which itself is a Gaussian, with mean and standard deviation
\begin{equation}
  \mu = \frac{\mu_1\sigma_2^2 + \mu_2\sigma_1^2}{\sigma_1^2 + \sigma_2^2}
\end{equation}
\begin{equation}
  \sigma = \frac{\sigma_1^2\sigma_2^2}{\sigma_1^2 + \sigma_2^2}.
\end{equation}

This sampling step has been implemented in \texttt{SLAVE} and we have
successfully tested it on simulated maps. With $N_{\textrm{side}}=512$ and
$l_{\textrm{max}}=1300$ and full sky coverage, we find $\alpha = 1.000 \pm
0.001$. The chains for the noise amplitude
$\alpha$ are shown in figure \ref{fig:alpha_simulated}.
Note that with such high resolution, the standard deviation on
$\alpha$ is extremely low, and any deviation from the exact $\alpha =
1.0$ will be detected. 

\section{Running SLAVE}
In this section, we quickly review how to use \texttt{SLAVE}. For a more
detailed usage, please see the \texttt{SLAVE} documentation (when the framework
will be released).

\texttt{SLAVE} requires the \texttt{HEALPIX} \citep{gorski:2005} CXX-libraries
installed. Please see the \texttt{HEALPIX} documentation on this topic. \texttt{SLAVE}
is run command-line, and requires a parameter file as command-line
parameter. The most important options in the parameter file are listed
in table \ref{tab:main}.

\begin{deluxetable*}{lll}
\tablewidth{0pt}
\tablecaption{\texttt{SLAVE} parameter table \label{tab:main}} 
\tablecomments{The \texttt{SLAVE} parameter names and usage may have changed
  when the first version is released. } 
\tablecolumns{3}
\startdata
\cutinhead{General parameters}
\texttt{seed}      & int  & Initial random seed \\
\texttt{verbosity}      & int  & Text output level (0=none) \\
\texttt{healpix\_dir}      & string  & HEALPIX home directory  \\
\texttt{output\_sigmas}      & bool  & Output $\sigma_\ell$ or not  \\
\texttt{output\_cls}      & bool  & Output $C_\ell$s or not  \\
\texttt{output\_directory}      & string  & Output file directory  \\
\texttt{output\_chisq}      & bool & Output the $\chi^2$ or not  \\
\texttt{output\_beam}      & bool  & Output the beam or not  \\
\texttt{output\_beam\_file}      & string  & Beam output filename  \\
\cutinhead{Operations}
\texttt{method}      & string  &  Analysis type: brute force\_fullsky or CG (normal)  \\
\texttt{CG\_convergence}      & double  & CG Convergence criteria (type $10^{-6}$)\\
\texttt{preconditioner}      & string  & Pre-conditioner type: none, static or 3j   \\
\texttt{init\_powerspectrum\_power}      & double  & Initialized flat power spectrum value  \\
\texttt{init\_powerspectrum\_use\_file}      & bool  & Use file instead of flat power spectrum  \\
\texttt{init\_powerspectrum\_file}      & string  & Initial power spectrum file \\
\texttt{samples}      & int  & Number of Gibbs samples to produce  \\
\texttt{burnin}      & int  & Number of burn-in samples to reject  \\
\cutinhead{Data}
\texttt{datasets}      & int  & Number of data sets (only 1 allowed yet..)  \\
\texttt{data\_nsideN}      & int  & $n_{\textrm{side}}$ for data set $N = \{1,2,3,\dots\}$   \\
\texttt{data\_mapN}      & string  & FITS map for data set $N = \{1,2,3,\dots\}$   \\
\texttt{data\_rmsN}      & string  & FITS rms map for data set $N = \{1,2,3,\dots\}$   \\
\texttt{data\_maskN}      & string  & FITS mask for data set $N = \{1,2,3,\dots\}$   \\
\texttt{beam\_fileN}      & string  & FITS beam for data set $N = \{1,2,3,\dots\}$   \\
\texttt{lmax}      & int  & $\ell_{\textrm{max}}$ for the analysis  \\
\texttt{constant\_rms}      & bool  & Use constant rms or not  \\
\texttt{constant\_rms\_value}      & double  & Value of constant rms  \\
\texttt{gaussian\_beam}      & bool  & Use a Gaussian beam or not  \\
\texttt{gaussian\_beam\_fwhm}      & double  & Value of Gaussian beam  \\
\cutinhead{Noise estimation parameters}
\texttt{enable\_noise\_amplitude\_sampling}      & bool & Enable noise estimation or not  \\
\texttt{noise\_sampling\_sigma}      & double  & The noise prior sigma  \\
\texttt{noise\_amplitude\_filename}      & string  & Output noise filename  \\
\texttt{noise\_alpha\_init\_val}      & double  & Initial value for $\alpha$  \\
\cutinhead{Binning}
\texttt{use\_binning}      & bool  & Enable binning of power spectrum  \\
\texttt{binning\_powerspectrum}      & string  & Power spectrum used for binning  \\
\texttt{bins\_filename}      & string & Text output the bins  
\enddata
\end{deluxetable*}

\subsection{Post-processing}
\begin{figure}
\mbox{\epsfig{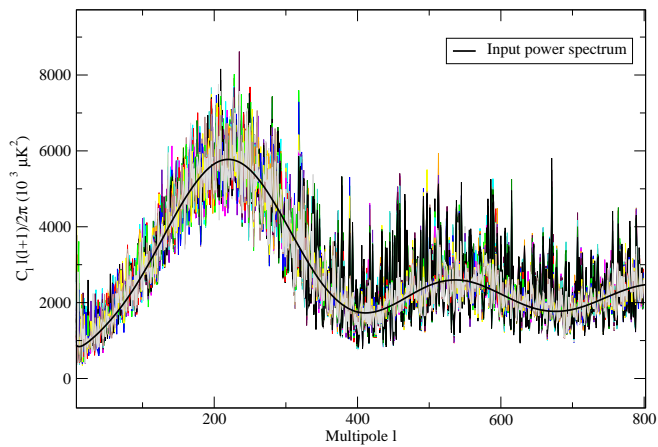}}
\caption{A typical plot of the $C_\ell$s obtained from a \texttt{SLAVE}
  run. Note that the input power spectrum is presented in black, and
  that the noise RMS for this particular run is very low. }
\label{fig:result_cls}
\end{figure}

After the Gibbs sampler has been cooking for a while, it is time to
investigate the results. The main output of \texttt{SLAVE} are the estimated power
spectra $C_\ell$'s and the signals $s$. However, as the signal is
assumed to be statistically isotropic, we instead output the signal power
spectra $\sigma_\ell$ defined as: 
\begin{equation}
  \sigma_\ell \equiv \frac{1}{2\ell +1} \sum_{m=-\ell}^{m=\ell}
  |s_{\ell m}|^2.
\end{equation}
The text-files may be plotted
directly through software such as \texttt{XMGRACE}, as presented in figure
\ref{fig:result_cls}.
In addition, \texttt{SLAVE} outputs the $\sigma_\ell$'s as a binary file for each chain. These binary
files can be combined through 
the main post-processing software utility for
\texttt{SLAVE} called \texttt{SLAVE\_PROCESS}. This software will combine the binary chains into a single
file, in addition to removing burn-in samples. To combine the sigmas into one file, type 
\begin{verbatim}
 slave_process 1 [no_chains] [no_samples] 
                 [burnin] [output sigma_l file]

\end{verbatim}

\begin{figure*}
\mbox{\epsfig{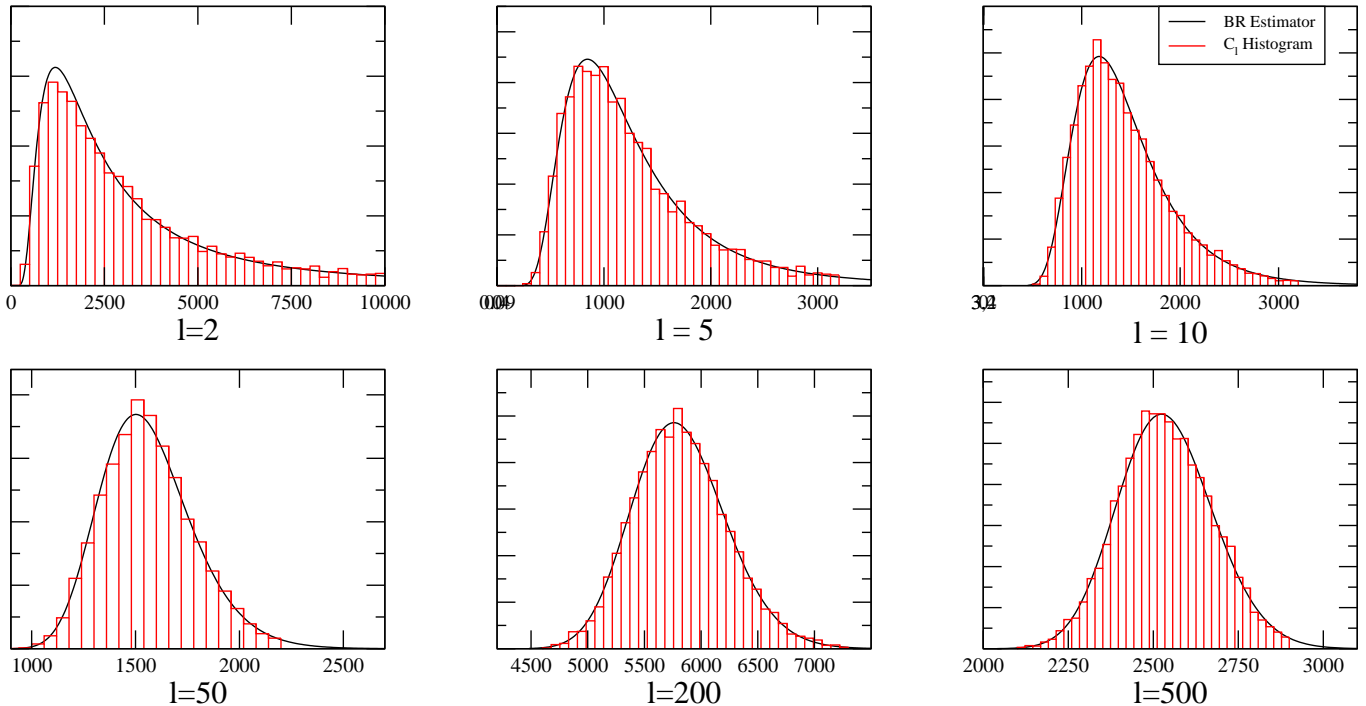}}
\caption{The histograms of the $C_\ell$s (red) and the BR-estimated
  likelihoods (black) for various $\ell$. Note how the distribution
  converges to a Gaussian for larger multipoles $\ell$. The analysis
  has been performed on simulated WMAP-like data.}
\label{fig:BR_likelihoods}
\end{figure*}

\subsection{$C_\ell$ likelihoods}
The first important step is to verify that the output $C_\ell$s follow
the desired inverse-Gamma distribution for low $\ell$, but converges to
Gaussians for larger $\ell$. The \texttt{SLAVE} processing utility
\texttt{SLAVE\_PROCESS} can generate a set of $C_\ell$s from the $\sigma_\ell$s
and output the corresponding values for a single $\ell$. It is then
straight-forward to use a graphical utility such as \texttt{XMGRACE} to obtain
the histogram. Such histograms are plotted together with the
analytical likelihoods in figure \ref{fig:BR_likelihoods}. Note the
good match between the histogram of the $C_\ell$s and the likelihoods
obtained from the Blackwell-Rao estimator. The analysis for producing
these plots was performed on simulated high-detail data, in order to
verify the validity of the BR-estimator.

To save the cls for a specific $\ell$, type
\begin{verbatim}
 ./process 4 [sigma_l file] [l] [generate no cls] 
             [output textfile] 
\end{verbatim}

\subsection{The Blackwell-Rao estimator}
Our primary objective is obtaining the best-fit power
spectrum from the estimated signal power spectra. If the $C_\ell$s were completely
distributed according to a Gaussian, one would only need to select the maximum of the
distribution for each $C_\ell$. However, as we saw in equation
\ref{eq:invgamma}, this is not the case, and we
need a better way to obtain the likelihood $\mathcal L(C_\ell)$ for each $\ell$. 
 
Luckily, we can obtain an analytical expression of the likelihood for
the $C_\ell$s via the Blackwell-Rao (BR) estimator, as
presented in \cite{chu:2005}. By using prior knowledge of the
distributions of the $C_\ell$s, we can build an analytical expression
for the distribution for each $C_\ell$ given the signal power spectrum
$\sigma_\ell$, or $P(C_\ell | \sigma_l)$.

Note that since the power spectrum only depends on the data through
the signal and thus $\sigma_\ell$, then  
\begin{equation}
  P(C_\ell \, | \, s,d) = P (C_\ell \,|\,s) = P(C_\ell \, | \, \sigma_\ell).
\end{equation}
It is therefore possible to approximate the distribution  $P(C_\ell \,
| \,d )$ as such:
\begin{eqnarray}
   P(C_\ell \, | \,d )  & = & \int P(C_\ell , s \, | \, d)\, ds  \\
                       & = & \int P(C_\ell \, | \, s,d) P(s\,|\,d) \,ds \\
                       & = & \int P(C_\ell \, | \, \sigma_\ell) P(\sigma_\ell\,|\,d) \,D\sigma_\ell \\ 
                       & \approx & \frac{1}{N_G} \sum_{i=1}^{N_G} P(C_\ell \, | \, \sigma_\ell^i)  
\end{eqnarray}
where $N_G$ is the number of Gibbs samples in the chain. This method of
estimating the $P(C_\ell \, | \,d )$ is called the Blackwell-Rao
estimator. Now, for a Gaussian field,

\begin{equation}
   P(C_\ell \, | \, \sigma_\ell )  \propto
   \prod_{\ell=0}^\infty \frac{1}{\sigma_\ell} \Big(
   \frac{\sigma_\ell}{C_\ell} \Big) e^{\frac{2\ell+1}{2}\frac{\sigma_\ell}{C_\ell}}.
\end{equation}
Taking the logarithm, we obtain a nice expression
\begin{equation}
\textrm{ln} P(C_\ell | \sigma_l) = \sum \Big( \frac{2\ell +1}{2}
\Big[-\frac{\sigma_\ell}{C_\ell} +  \textrm{ln} \big( \frac{\sigma_\ell}{C_\ell}
\big) \Big] - \textrm{ln} \sigma_l \Big)
\end{equation}
which is straight-forward to implement numerically. 
To output the BR-estimated likelihood for one $\ell$, type 
\begin{verbatim}
 ./process 3 [sigma_l file] [l] 
             [output likelihood]
\end{verbatim}

\subsection{Power spectrum estimation}
\begin{figure}
\mbox{\epsfig{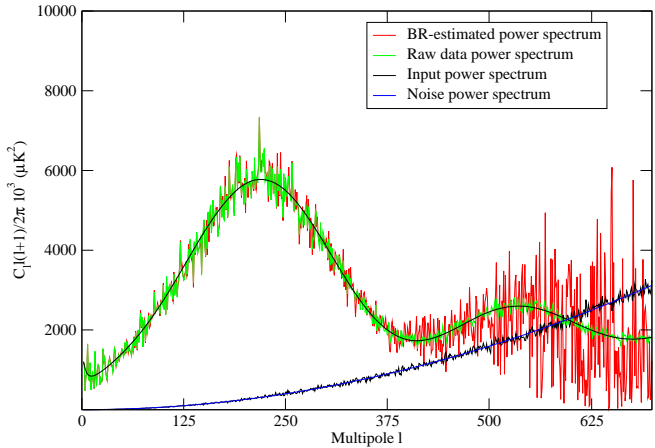}}
\caption{The BR-estimated power spectrum (red) versus the simulated input data
  power spectrum (green). Note that these two power spectra agree on
  large scales. The noise power spectrum is also shown (blue).}
\label{fig:BR_powerspectrum}
\end{figure}

The best-fit BR-estimated power spectrum is obtained by choosing
the maximum likelihood value of $C_\ell$ for each $\ell$. To do so,
type
\begin{verbatim}
 ./process 2 [sigma_l file] 
             [output power spectrum file]
\end{verbatim}
An example of a BR-estimated power spectrum can be seen in figure
\ref{fig:BR_powerspectrum}. In addition, both the input-and noise
power spectra are shown. Note how the BR-estimated power spectrum is
exact on small scales (low $\ell$), while the convolution and noise
dominated on higher scales. 

\section{Conclusions}
\label{sec:results}
We have presented a self-contained guide to a CMB Gibbs sampler,
having focused on both deriving the conditional probability distributions
and code design. We described in detail how one can draw
samples from the conditional distributions, and saw how the Gibbs
sampler is numerically superior to conventional MCMC methods, scaling
as $\mathcal O(n^{1.5})$.  We have also introduced a new object-oriented CMB Gibbs
framework, which employs the existing \texttt{HEALPix} \citep{gorski:2005} C++
package. We presented a small guide to the usage of \texttt{SLAVE}, including
post-processing tools and the Blackwell-Rao estimator for obtaining
the likelihoods and the best-fit power spectrum. We also reviewed a new way
of estimating noise levels in CMB maps, as presented in
\cite{groeneboom:2009a}. The software package \texttt{SLAVE} will
hopefully be released when it is completed during 2009, and will
run on all operating systems supporting the GNU C++ compiler. Please
see \texttt{http://www.irio.co.uk} for release details and information.

\begin{acknowledgements}
  Nicolaas E. Groeneboom acknowledges financial support from the Research Council
  of Norway. Nicolaas especially wishes to thank Hans Kristian
  Eriksen, but also Jeffrey
  Jewell, Kris Gorski, Benjamin Wandelt and the whole ``Gibbs team''
  at Jet Propulsion Laboratories (JPL) 
  for useful discussions, comments and input.  
  The computations presented in this paper were carried out
  on Titan, a cluster owned and maintained by the University of Oslo
  and NOTUR. We acknowledge use of the
  \texttt{HEALPix} \footnote{http://healpix.jpl.nasa.gov} software
  \citep{gorski:2005} and analysis package for deriving the results in
  this paper. We acknowledge the use of the Legacy Archive for
  Microwave Background Data Analysis (LAMBDA). Support for LAMBDA is
  provided by the NASA Office of Space Science. 
\end{acknowledgements}

\clearpage

\clearpage

\end{document}